\input{aipcheck}

\documentclass[
    ,final            % use final for the camera ready runs
  ]
  {aipproc}

\layoutstyle{6x9}

\begin{document}

\title{Unsolved Problems about Supernovae}


\author{Nino Panagia}{
  address={Space Telescope Science Institute,3700 San Martin Drive, 
  Baltimore, MD 21218, USA }
  ,altaddress={INAF/Osservatorio Astrofisico di Catania, Via S.Sofia 78, 
  I-95123 Catania, Italy} % additional visiting address
  ,altaddress={Supernova Ltd., Olde Yard Village \#131, Northsound Road, 
  Virgin Gorda, British Virgin Islands} 
  ,altaddress= {ICRANet, Piazzale della Repubblica 10, I-65100 Pescara, Italy} % additional visiting address
}

\begin{abstract}
 A number of unsolved problems and open questions about the nature and
the properties of supernovae are identified and briefly discussed.  Some
suggestions and directions toward possible solutions are also
considered.

\end{abstract}

\maketitle

%%%%%%%%%%%%%%%%%%%%%%%%%%%%%%%%%%%%%%%%%%%%
%% MAINMATTER
%%%%%%%%%%%%%%%%%%%%%%%%%%%%%%%%%%%%%%%%%%%%

\section{Introduction}

  When I started studying supernovae (SNe) in 1978, I thought that most
problems had been solved already and that I could just help clarifying
minute details.  Actually, there were quite a number of fundamental
aspects still to understand and clarify. In the course of many years, I
have done my share of work, e.g. establishing multifrequency studies of
SNe to clarify the nature of the overall explosion process among the
different classes \cite{pan80}, starting systematic radio studies of SNe
that enable one to probe the circumstellar medium of the SN progenitors
\cite{wei81},  recognizing the existence of different types of type I
SNe, which indicated that the absence of hydrogen in their spectra was
not telling us the full story \cite{pan85},  finding observational
evidence that  SNIa are produced by at least two different channels of
progenitors \cite{dp03,detal05,manetal05,mdp06}, and, maybe,  a few more
items.  However, these are a minor part of the problems successfully
addressed in the last 30 years, and a tiny number as compared with  the
ones that still need a solution.

In this talk I will consider a few of the hottest or most intriguing
problems yet to solve about the nature and the properties of
supernovae. 

\section{SNIa Progenitors and Properties}

\subsection{SNIa Progenitors: Why Binary Systems?}

Common wisdom has it that SNIa originate from moderate mass binary
systems (e.g. \cite{nometal97}), but what is the
evidence for binarity? Let's see:\\
- The absence of H and He from SNIa ejecta points in the direction of
highly evolved progenitors. \\
- Judging from SNIa rates in various types of galaxies (e.g.
\cite{mdp06,suletal06}),  about 50\% of 
SNIa explosions occur hundreds million or even several billion
years after the formation of the progenitor star. \\
- It is an experimental fact that Nova explosions are produced in binary
system in which matter accreted on a white dwarf from its close
companion induces sudden nucleosynthesis in the superficial layer of accreted
material.\\
- In order to produce an energetic explosion (say, $>10^{51}$ erg) the
explosive nuclear burning of a substantial amount of mass (say,
$>1M_\odot$) is required, and, therefore,  the progenitor star must have
mass substantially higher than the Sun and suitably lower than what is
believed to give rise to core-collapse explosions, about 8M$_\odot$.\\ -
Invoking a binary system increases dramatically the physical channels
capable to produce a SNIa explosion (through the introduction of more
free parameters...)  and may help justifying the wide range of
properties of different SNIa (e.g. Benetti et al. \cite{benetal05}).

The fact is that we have never seen any such star actually exploding,
nor can we expect a detection any time soon because white dwarfs are
faint stars and even their companions are not expected to be
particularly bright stars.  Thus, we cannot identify the progenitor
stars before their explosions (e.g. on historical pre-SN images) nor it
is going to be easy to detect a possibly surviving companion. We are
left with circumstantial evidence, but we have to make it as solid as
possible before concluding that {\it "we know what the progenitors
are"} (e.g. Phillips \cite{phil93}, and this Conference).

\subsection{Stellar Evolution of SNIa Progenitors}

As mentioned before, it is most reasonable that SNIa progenitors are
stars in the approximate range of 3-8 M$_\odot$, i.e.  stars that
on the Main Sequence belong to spectral types B.  Many of these stars are known
to display high rotational velocities, and may be expected to have high
magnetic fields as a result of intense dynamo activity.  Perhaps some of
these properties may be THE cause for a given B-type star to end its
evolution exploding as a SNIa, whether with or without the help of a
companion.

Moreover, stars in the mass range 3-8 M$_\odot$ are known to become Cepheid
variables when they evolve off the main sequence and to end their {\it
regular} evolution through a Planetary Nebula phase.  
Perhaps there are important clues to the nature of SNIa progenitors that
can be extracted from the study of stars easily identifiable in these
characteristic phases.   For example, a systematic study of the
white dwarf companions of Cepheid variables would permit us to clarify
the statistical properties of binary systems that may give origin to a
SNIa and help identifying the crucial parameters that lead to a SNIa
explosion.

\subsection{SNIa as Standard Candles for Cosmology}

Even if SNIa are commonly referred to as {\it ideal standard candles}
because of their high luminosity at explosion and their relatively
narrow range of observational properties, they are not quite {\it
perfect} and, maybe, {\it not even standard}.  Actually, it is widely
recognized that their peak luminosities span more than a factor of ten
and that their spectral properties do not correlate closely with their
photometric properties. Standardization is obtained exploiting  the
empirical evidence that bright SNIa evolve photometrically more slowly
than faint SNIa. However, even if this phenomenon is measured rather
accurately in the local Universe so as to lead to accurate empirical
corrections, its physical nature is far from understood and, therefore,
this standardization rests on unsecure grounds.  Moreover,  when moving
to higher redshifts \cite{mdp06} , i.e. to the past phases of the
evolution of the Universe, the balance between fast evolving SNIa
progenitors (with lifetimes less than about 100 Myrs) and the ``tardy"
ones (that typically explode more than 1 Gyr after formation) shifts in
favor of the fast component, so that any systematic difference between
the two components will bias the observational results.

A good summary of the current efforts to address and solve these
problems can be found in Andy Howell et al. Astro-2010 White Paper on
SNIa \cite{howetal09}.

\section{ Core Collapse Supernova Progenitors}

Core-collapse supernovae (CC-SNe) are believed to originate from the
collapse of the core of massive stars (M$>8$M$_\odot$) at the end of
their nucleosynthetic life when Iron has formed in the core of the star
and there is no additional energy source to prevent the star from
collapsing onto itself catastrophically. 

Evidence that this is the correct scenario for SNe in this class has
been provided by the detection of a total production of  more than
$10^{53}$ erg in the form of neutrinos from SN 1987A explosion: this was
just what theoretical predictions were expecting from the collapse of a
massive stellar core (e.g. Arnett et al. 1989 \cite{arnetal89}).

However, SN 1987A was exceptional in that it occurred so close to us (a
mere 52kpc...) that it was easy to identify the star that exploded and
monitor what was left behind after the explosion.  At distances of
several Mpcs or more it is much harder to separate stars in compact
clusters (for example, at 5Mpc even the 0.05" resolution of HST optical
imagers corresponds to an absolute size of 1.2pc) and the identification
is necessarily of statistical value and can be affected by large
uncertainties.

\subsection {Direct Identification of Progenitors}

Nevertheless, when something is hard, it is still possible... And with a
judicious/clever observing strategy one can obtain the necessary
information on the  SN progenitors.  Thus, comparing post-explosion high
resolution (essentially HST) images with pre-explosion images of the
parent galaxies it is possible to recognize which star did explode. And
if one has images taken with different filters one can determine the
stellar properties of the progenitor.

In this way, the heroic efforts of the Belfast Group (Steve Smartt and
collaborators) and California Group (Schuyler Van Dyk, Avishai Gal-Yam
and collaborators) have provided clear measurements of the properties of
Type IIP SN progenitors, leading to the identification of red supergiant
progenitors and to explicit mass determinations for about a dozen SNe,
which fall in the range $\sim8$M$_\odot$ to $\sim16$M$_\odot$.

Trying to do the same for SNIb/c has not been as successful, mostly
because these events are more rare and, therefore, are likely to be
discovered in more distant galaxies so that even HST resolution is not
enough to provide an unambiguous identification of the progenitor star. 
In the cases in which a positive identification has been possible, the
progenitors seemed to be  early type stars, as expected in the scenario
that calls for WR stars to produce SNIb/c.  However, just because of
being high effective temperature stars, photometric measurements in few
optical bands do not allow one to determine pre-explosion bolometric
luminosities and/or stellar masses.  The few lower limits 
indicate that the targeted SNIb/c had progenitors with masses
above
$\sim$20 M$_\odot$.

While these results seem to be consistent with theoretical and
statistical expectations, one has to consider that a lower limit is
hardly good enough to clarify the issues but can at best reduce the
parameter space within which one has to look for THE correct answer.

\subsection{ The SNII zoo: What are SNIIb, SIIn, SNIIL?}

Still we don't know much about the remaining multiplicity of CC-SNe,
such as SNIIb, SIIn, SNIIL.  The progenitor mass estimates for SNIIP and
SNIb/c leaves unanswered the question "are the progenitors masses of 
SNIIn, SNIIL, SNIIb falling in the apparent gap 16-20 M$_\odot$ or are
these SN explosions mostly  determined by some other parameter, such as
binarity, rotation, etc?"  As usual, it looks as if ALL of
these parameters may enter crucially into the game: for instance, radio
observations have shown that the SNIIL 1979C exploded in a detached
binary system with a primary star around 20 M$_\odot$ and a companion of
at least 5 M$_\odot$ \cite{wetal92}, SNIIb 2001ig  displayed radio light
curves highly suggestive of interactive winds as seem in systems
composed of massive early stars, and possibly containing a WR star
\cite{rydetal04}, and the progenitors of  both  SNIIb 1993J and SN200igd
had mass loss rates that were strongly decreasing in the last 10,000
years before their explosions \cite{stocketal07,wetal07}.

\subsection {Ultra-faint Supernovae}

While the luminosities of most type Ia SNe are believed to be confined in a
relatively  narrow interval, it is well known that the luminosities of CC-SNe
can vary by large factors, more than a factor of 100.  It is intuitive
that if there is a category of faint CC-SNe, one may miss many, or most
of them unless looking very hard.  The studies championed by Benetti, 
Pastorello, and their group have shown that some SNII can be really
faint, as faint as $M_B\simeq-14$  and perhaps even fainter \cite{pastetal07}. Despite
their valuable efforts and results (many of them presented at this
Conference), we are seeking quantitative answers the the questions: {\it
``How faint can ``faint" SNe be?"},  {\it ``Where is the boundary
between Nova and Supernova phenomena fall?"},  {\it ``How many ``faint"
SNe are there?"}, and {\it ``How many ``unexpected SN explosion channels
exist?"}

\subsection {Ultra-bright Supernovae}

At the other end of the scale, it is clear that exceptionally bright
SNII do occur, starting with ``my" first SN, 1979C, which at maximum was
at least as bright as a SNIa ($M_B<-19.5$) \cite{pan80}, to the very recent SN
2005ap \cite{quetal07} and 2006gy \cite{agnetal09}.  These events pose a number of question, such as
{\it ``How bright can ``ultra-bright" SNe be?"}, {\it ``What are
they?"}, {\it ``Why do they occur?"}, etc.  One of the ``simple"
suggestions is that they are the result of a pair-annihilation event
which is predicted for ultramassive stars (M$>140$M$_\odot$) of
extremely low metallicities (Z$<1/1000$Z$_\odot$).  This is an
interesting possibility that is worth considering and  pursuing, but I
find it hard to imagine that such stars may exist in the local Universe,
after more than 13 billion years of chemical evolution.

\subsection{The end of the most massive stars}

Another open problem is what happens to the most massive stars,
i.e. those stars with initial masses in the approximate range 30-40-up
to 100M$_\odot$.  They are supposed to end their lives as black holes
and, possibly, end in a quite unspectacular event.  So, one is presented
with the questions:  {\it ``Do these stars explode producing very  faint
events or do they just ``disappear"?}, and  {\it ``Can we detect such
events?"}. One may think that detecting a "non-explosion" (i.e. the lack
of an explosion)  is a self-contradictory statement because it is the
same as detecting ``nothing".  However, even in the case of a true
"non-explosion" there are ways to notice such an event because the star
that undergoes a quiet end into its ``final nothing", by virtue of being
quite massive, was a bright star {\it before} its end and suddenly it
drops in luminosity, essentially dropping forever from sight. 
Therefore, one should monitor at regular intervals a suitably large
number of reasonably nearby galaxies and record which bright stars have
disappeared. This is exactly what the project lead by Chris Kochanek
\cite{koetal08} is doing and, possibly,  in a few years we will have some
exciting news about the demise of really massive stars.

\section {The GRB-SN Connection}

It is now well established that a number of Gamma-Ray Burst (GRB)
sources have associated SNIc counterparts (e.g. \cite{dlv07}) and that
both SNIc and GRBs are most closely associated with active star forming
regions than any other type of explosive sources \cite{fretal06}.

However, only a handful of GRBs-SNIc associations have been recognized
so far
and, despite all efforts, there are GRBs that are positively not
associated with any SN, e.g. GRB 060614 for which it has been excluded
the association with any SN brighter than -13.7 
\cite{detal06,galetal06}, and many SNIc that have not shown evidence of
any conspicuous gamma-ray emission (e.g. \cite{dlv07}).

Therefore, we are presented with two complementary and fundamental
questions:  {\it ``Why many SNe are not GRBs?"}, and  {\it ``Why many
GRBs are not SNe?"}.  Only answering these questions, we could claim
that we do understand the GRB-SN connection.

\section {Conclusions and Recommendations}

My main conclusion from this series of considerations is that there is
still a lot to learn  about the properties and the nature of most types
of  SNe, but it appears that we (actually... YOU!) are working hard at
it! All the interesting presentations of this week have shown how
seriously the existing problems are been addressed, and I am confident 
that with hard work and perseverance, and a pinch of luck, we will
eventually {\it ``see the light at the end of the tunnel"}.

Still, I have a few suggestions and recommendations to make.

First, it seems to me that, since all SNe appear to originate from
progenitors with original masses above 3 M$_\odot$, i.e. B-type or
O-type stars, both rotation and magnetic fields can be crucially
important and should explicitely be folded in the theorical model
evolution 
that  leads to a SN explosion.

While studying  any type of SN and any aspect of their explosions, one 
should also consider  the cosmic evolution of the SN phenomenon as a
function of important parameters such as  {\it (i)} metallicity, {\it
(ii)} stellar
population, {\it (iii)} cosmic age, and {\it (iv)} environment.  With good
imagination and a sharp mind you can add here more relevant parameters,
but, {\it please}, do not neglect considering  any of the ones listed
here.  The point is that we are not studying SNe just to understand SNe,
but rather because we want to understand how the Universe works and
evolve.

In pursuing these noble goals, I have a recommendation for  everyone and,
especially for the younger generations: {\it ``When aiming at solving a
problem try to  explore ALL possible solutions"}, or {\it ``Do NOT stop
at the first possible solution and call it THE solution"}. Here the
point is that  declaring success after finding a solution may help along
your path to fame (or to tenure...) but could also be a terrible
disservice for science if the solution is not unique!

Also, I recommend that the SN community should  devise (and enforce...)
a classification that is based on Physics rather than pure morphology. 
Defining new subclasses and/or catchy names every time something
different is found may not help much to the understanding of the complex
phenomen of SN explosions, and may slow down the progress in some
fields by creating the illusion that a problem has been solved.

Finally, I believe that all SN data should be collected in   a central
database that is properly arranged and open to everyone to search and
peruse.  A general archive that contains all the observational data
would allow one to study all possible connections in a systematic way
and compare the results with existing models and theories.  Only by
accounting for {\it ALL} aspects of a problem can one claim to have
found {\it THE} solution.

%%%%%%%%%%%%%%%%%%%%%%%%%%%%%%%%%%%%%%%%%%%%%%%%
%% BACKMATTER
%%%%%%%%%%%%%%%%%%%%%%%%%%%%%%%%%%%%%%%%%%%%%%%%

\begin{theacknowledgments}
This work was supported in part by STScI-DDRF Grant D0001.82392.  I wish
to thank Massimo Della Valle, Filippo Mannucci and Kurt Weiler for
valuable  discussions on various aspects of SN studies, and the
organizers of this Conference for giving me an opportunity to
participate in it.
\end{theacknowledgments}
%%%%%%%%%%%%%%%%%%%%%%%%%%%%%%%%%%%%%%%%%%%%%%%%
%% The bibliography can be prepared using the BibTeX program or
%% manually.
%%
%% The code below assumes that BibTeX is used.  If the bibliography is
%% produced without BibTeX comment out the following lines and see the
%% aipguide.pdf for further information.
%%
%% For your convenience a manually coded example is appended
%% after the \end{document}
%%%%%%%%%%%%%%%%%%%%%%%%%%%%%%%%%%%%%%%%%%%%%%%%

%%%%%%%%%%%%%%%%%%%%%%%%%%%%%%%%%%%%%%%%%%%%%%%%
%% You may have to change the BibTeX style below, depending on your
%% setup or preferences.
%%
%%
%% For The AIP proceedings layouts use either
%%%%%%%%%%%%%%%%%%%%%%%%%%%%%%%%%%%%%%%%%%%%

\bibliographystyle{aipproc}   % if natbib is available
%\bibliographystyle{aipprocl} % if natbib is missing

%%%%%%%%%%%%%%%%%%%%%%%%%%%%%%%%%%%%%%%%%%%
%% You probably want to use your own bibtex database here
%%%%%%%%%%%%%%%%%%%%%%%%%%%%%%%%%%%%%%%%%%%
%\bibliography{sample}

%%%%%%%%%%%%%%%%%%%%%%%%%%%%%%%%%%%%%%%%%%%
%% Just a reminder that you may have to run bibtex
%% All of it up to \end{document} can be removed
%% if you don't like the warning.
%%%%%%%%%%%%%%%%%%%%%%%%%%%%%%%%%%%%%%%%%%%

%%%%%%%%%%%%%%%%%%%%%%%%%%%%%%%%%%%%%%%%%%%
%% The following lines show an example how to produce a bibliography
%% without the help of the BibTeX program. This could be used instead
%% of the above.
%%%%%%%%%%%%%%%%%%%%%%%%%%%%%%%%%%%%%%%%%%%

\end{document}